\documentclass[letterpaper]{ptephy_v1}

\preprintnumber{XXXX-XXXX} 

\usepackage{url} 

\begin{document}

\title{First application of superconducting transition-edge-sensor
microcalorimeters to hadronic-atom x-ray spectroscopy}

\newcommand{\RIKEN}{1}
\newcommand{\NIST}{2}
\newcommand{\LNF}{3}
\newcommand{\LUND}{4}
\newcommand{\UT}{5}
\newcommand{\NARAWU}{6}
\newcommand{\TOTTORI}{7}
\newcommand{\KEK}{8}
\newcommand{\OSAKAC}{9}
\newcommand{\TMU}{10}
\newcommand{\SMI}{11}
\newcommand{\OSHIMA}{12}
\author[\RIKEN,*]{\collaborator{HEATES Collaboration}S. Okada\thanks{Present address: AMO physics laboratory, RIKEN, Wako, 351-0198, Japan}}

\affil[\RIKEN]{RIKEN Nishina Center, RIKEN, Wako, 351-0198, Japan}
\affil[\NIST]{National Institute of Standards and Technology, Boulder, CO 80305, USA}
\affil[\LNF]{Laboratori Nazionali di Frascati dell' INFN, I-00044 Frascati, Italy}
\affil[\LUND]{Lund University, Box 117, 221 00 Lund, Sweden}
\affil[\UT]{Department of Physics, The University of Tokyo, Tokyo, 113-0033, Japan}
\affil[\NARAWU]{Department of Physics, Nara Women's University, Kita-Uoya Nishimachi, Nara 630-8506, Japan}
\affil[\TOTTORI]{Department of Regional Environment, Tottori University, Tottori 680-8551, Japan}
\affil[\KEK]{High Energy Accelerator Research Organization, KEK, Tsukuba, 305-0801, Japan}
\affil[\OSAKAC]{Research Center for Physics and Mathematics, Osaka Electro-Communication University, Neyagawa, Osaka 572-8530, Japan}
\affil[\TMU]{Tokyo Metropolitan University, Hachioji, 192-0397, Japan}
\affil[\SMI]{Stefan-Meyer-Institut f$\ddot{\mbox{u}}$r Subatomare Physik, A-1090 Vienna, Austria}
\affil[\OSHIMA]{National Institute of Technology, Oshima College, Oshima, Yamaguchi, 742-2193, Japan \email{sokada@riken.jp}}
\author[\NIST]{D.~A.~Bennett}
\author[\LNF]{C.~Curceanu}
\author[\NIST]{W.~B.~Doriese}
\author[\NIST]{J.~W.~Fowler}
\author[\NIST]{J.~Gard}
\author[\LUND]{F.~P.~Gustafsson}
\author[\RIKEN]{T.~Hashimoto}
\author[\UT]{R.~S.~Hayano}
\author[\NARAWU]{S.~Hirenzaki}
\author[\NIST]{J.~P.~Hays-Wehle}
\author[\NIST]{G.~C.~Hilton}
\author[\TOTTORI]{N.~Ikeno}
\author[\LNF]{M.~Iliescu}
\author[\KEK]{S.~Ishimoto}
\author[\RIKEN]{K.~Itahashi}
\author[\RIKEN]{M.~Iwasaki}
\author[\OSAKAC]{T.~Koike}
\author[\TMU]{K.~Kuwabara}
\author[\RIKEN]{Y.~Ma}
\author[\SMI]{J.~Marton}
\author[\RIKEN]{H.~Noda\thanks{Present address: Frontier Research Institute for Interdisciplinary Sciences, Tohoku University, Sendai, Miyagi 980-8578, Japan}}
\author[\NIST]{G.~C.~O'Neil}
\author[\RIKEN]{H.~Outa}
\author[\NIST]{C.~D.~Reintsema}
\author[\RIKEN]{M.~Sato}
\author[\NIST]{D.~R.~Schmidt}
\author[\LNF]{H.~Shi}
\author[\SMI]{K.~Suzuki}
\author[\UT]{T.~Suzuki}
\author[\NIST]{D.~S.~Swetz}
\author[\KEK,\NIST]{H.~Tatsuno\thanks{Present address: Lund University, Box 117, 221 00 Lund, Sweden}}
\author[\LUND]{J.~Uhlig}
\author[\NIST]{J.~N.~Ullom}
\author[\SMI]{E.~Widmann}
\author[\TMU]{S.~Yamada}
\author[\OSHIMA]{J.~Yamagata-Sekihara}
\author[\SMI]{J.~Zmeskal}

\begin{abstract}
 High-resolution pionic-atom x-ray spectroscopy was performed
 with an x-ray spectrometer based on a 240-pixel array of
 superconducting transition-edge-sensor (TES) microcalorimeters
 at the $\pi$M1 beam line of the Paul Scherrer Institute.
 X-rays emitted by pionic carbon via the $4f \to 3d$ transition
 and the parallel $4d \to 3p$ transition were observed
 with a full-width-at-half-maximum energy resolution of 6.8 eV at 6.4 keV.
 Measured x-ray energies are consistent with
 calculated electromagnetic values which considered the strong-interaction effect
 assessed via the Seki-Masutani potential for the $3p$ energy level,
 and favor the electronic population
 of two filled $1s$ electrons in the K-shell.
 Absolute energy calibration with an uncertainty of 0.1 eV
 was demonstrated 
 under a high-rate hadron beam condition of 1.45 MHz.
 This is the first application of a TES spectrometer
 to hadronic-atom x-ray spectroscopy
 and is an important milestone towards next-generation high-resolution
 kaonic-atom x-ray spectroscopy.
\end{abstract}

\subjectindex{C30, D15, H15}

\maketitle

\section{Introduction}

A hadronic atom consists of a negatively-charged hadron
(e.g., $\pi^-$, $K^-$, $\bar{p}$, $\Sigma^-$, $\Xi^-$)
and electrons that are bound by a Coulomb field to an atomic nucleus.
Such a system can be used to probe the strong interaction
between hadrons and atomic nuclei in the low-energy limit.
The mass of the hadron, 
significantly larger than the mass of the electron it replaces,
shifts the atomic transition energies in the hadronic atom to
significantly higher energies than those of the standard atom.
In addition, effects of the strong interaction
appear in the most tightly bound energy level.
These perturbations include an energy shift from that given only by the
electromagnetic interaction, and a lifetime broadening 
due to absorption of the hadron by the nucleus.
The shift and width can be experimentally
extracted via x-ray-emission spectroscopy
of characteristic transitions into this lowest orbital.

In kaonic atoms,
the understanding of the low-energy $\overline{K} N$ interaction
has been substantially deepened
by the most recent kaonic-hydrogen-atom measurement \cite{SidKp}
and theoretical studies (e.g., \cite{Ikeda}).
However, the depth of the $\overline{K}$-nucleus potential
remains unknown
because of insufficient precision in the kaonic-atom data for $Z\ge2$;
this is one of the greatest present concerns
in strangeness nuclear physics \cite{Gal13}.

In recent years, Balmer-series x-rays of the $K^-$-He atom ($\sim$ 6 keV)
were measured at the K5 beamline of KEK-PS \cite{E570}
and the DA$\Phi$NE electron-positron collider
of the Laboratori Nazionali di Frascati \cite{SidHe4,SidHe3}.
These measurements followed discussions of the importance of the measurement of 
the strong-interaction shift and width of the $2p$ level
in kaonic-helium atoms ($K^-$-He) over 30 years ago \cite{Bai83,Aka05}.
Both experiments employed Silicon Drift Detectors (SDDs)
whose FWHM energy resolution is typically $\sim$ 200 eV
at 6 keV \cite{E570,SidHe4,SidHe3}.
To observe spectral effects
as predicted by theoretical calculations \cite{Hir00,Bat90,Fri11},
e.g., a 0.2 eV shift and 2 eV line width,
a new approach with a significantly improved resolution is essential.

We are preparing a high-resolution x-ray measurement
of kaonic atoms at a kaon beamline of J-PARC \cite{K1.8BR}
using a novel superconducting transition-edge-sensor (TES)
microcalorimeter (J-PARC E62 \cite{E62}).
The FWHM energy resolution of this type of detector can be
as good as about 2 eV at 6 keV \cite{Smith12,Uhlig15}
which is about two orders of magnitude better than that of SDDs.
The spectrometer is a highly sensitive thermal sensor
that measures energy deposition
via the increase in the resistance
of a superconducting thin film that is biased
within the sharp phase transition
between the normal and superconducting phases.
The detailed working principles and the recent progress
of the TES system are described in Refs. \cite{Ens05,Ull15,Uhlig15}.

An alternate high-resolution x-ray spectroscopy technology
is based on diffraction from Bragg crystals.
However, they have not been used to study the strong interaction
in most hadronic atoms due to their low efficiency.
Only pion beams and the resulting x-ray emission are
intense enough to enable the study of pionic atom x-rays \cite{Mar76,Cha85,Sig95}                                                                                                                                                                                                                                                                                                                                                                                                                        
with crystal spectrometers.
On the other hand,
recent technological advances in multiplexed readout 
of multi-pixel TES arrays
(more than 100 pixels; each having $\sim$0.1 mm$^2$ effective area)
allow the performance of a precision kaonic-atom measurement
in a realistic data acquisition time.

Until this experiment, a TES x-ray spectrometer has never been utilized
in a hadron-beam environment.
To study the in-beam performance of the TES spectrometer
at a hadron beamline and demonstrate the feasibility
of TES-based hadronic-atom x-ray spectroscopy,
we performed a pioneering experiment at a pion beamline
by measuring the x-rays from the $4f \to 3d$ transition of
pionic carbon ($\pi$-$^{12}$C).
This x-ray transition was chosen because the energy ($\sim$ 6.5 keV)
is similar to the $K^-$-He $3 \to 2$ x-ray energy,
while the strong-interaction effects are negligibly small.

\section{Experiment}

\begin{figure}[t] 
\centering\includegraphics[width=1.\linewidth]{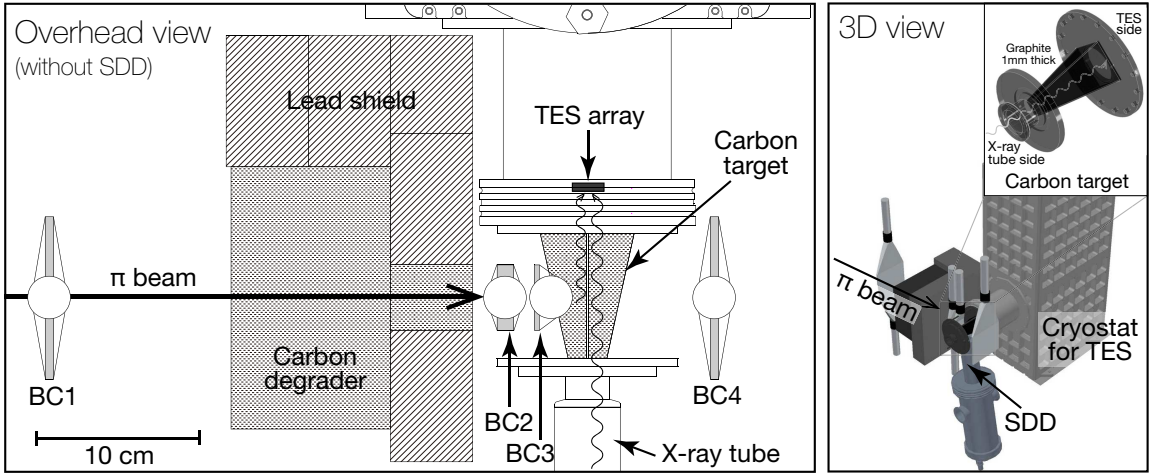}
\caption{A top view (left) and birds-eye view (right)
 of the experimental setup.}
\label{fig_setup}
\end{figure}

The experiment was carried out at the $\pi$M1 beamline \cite{piM1}
of the Paul Scherrer Institute (PSI) in October 2014.
Pions were created by a proton beam of energy 590 MeV at a current
of up to 2.3 mA passing through a graphite target located 
in the main beamline of the PSI synchro-cyclotron, 
and were transported through the $\pi$M1 beamline.

A schematic view of the experimental setup is shown in
Fig. \ref{fig_setup}.
With a 2.2 mA primary proton beam,
the $\pi^-$ beam rate with momentum 173 MeV/$c$
was 1.45 $\times$ 10$^6$ /sec
at BC1 (10$\times$10 cm$^2$ plastic scintillation counter)
located at a distance 30 cm upstream from the focal point.
The $\pi^-$/$e^-$ ratio was $\sim$0.7
due to a missing electrostatic separator in the beamline.

The incident $\pi^-$'s were degraded in moderators, 
counted with beamline counters (BCs),
and stopped in a carbon graphite target
where $\pi$-$^{12}$C atoms were generated.
X-rays were emitted by the highly excited atoms
and detected by a TES x-ray spectrometer
through a 150$\mu$m-thick beryllium vacuum window
and three layers of 5-$\mu$m-thick aluminum IR-blocking filters
(one at each of the three temperature stages: 50K, 3K, 50 mK).

A 240-pixel TES array \cite{Ull14} was the x-ray spectrometer.
Each pixel had a 4-$\mu$m-thick bismuth absorber 
(80\% absorption efficiency for 6.4 keV x-rays)
which converted the incident x-ray energy to heat;
the absorber was coupled to a sensitive thermal sensor, the TES,
composed of a superconducting Mo/Cu proximity bilayer film.
Each absorber had a collimated effective area 
of 320 $\mu$m $\times$ 305 $\mu$m,
thus the total collecting area of the array was about 23 mm$^2$.
Each TES pixel was electrically biased
to the superconducting critical temperature ($T_\mathrm{C} \sim$ 107 mK)
and to about 20 - 30\% of their normal resistance of $\sim$10 m$\Omega$.
A pulse-tube-backed adiabatic 
demagnetization refrigerator (ADR) \cite{HPD}
cooled the system to a regulated bath temperature 
of 75 mK $\pm$ 7 $\mu$K (rms).
The regulated hold time was about 36 hours,
after which the ADR cycle
(magnetic field increased isothermally and decreased adiabatically) took 2 hours
before another 36 hours of operation at 75 mK.

SQUID current amplifiers were used to read out the current signal
from the low-resistance TESs.
For the 240-pixel readout,
a time division SQUID multiplexing (TDM) scheme \cite{TDM} was employed
to reduce the number of wires running to the low temperature stages 
of the cryocooler.
The signal from each TES channel was coupled to a SQUID amplifier, 
and the outputs from 30 individual channels were switched sequentially 
and read out by a single amplifier. 
The multiplexing frame time was 9.6 $\mu$s (0.32 $\mu$s/channel).
The sampling rate was therefore 104 kHz for each pixel.
The 240 pixel readout was realized by use of eight TDM columns in parallel.

TES data were continuously streamed into a PC server,
and were recorded to disk 
only when a current ``pulse'' due to an x-ray event was triggered.
The system recorded 1024 samples (9.83 ms) for each event,
where the first 256 samples corresponded to a pre-triggered timing region.
The typical exponential rise and decay time constants of an x-ray pulse were
$\sim$200 $\mu$sec and $\sim$500 $\mu$sec respectively.
The x-ray energy corresponding to each pulse height was calculated
with an optimal-filter technique \cite{OptimalFilter}.
The data-acquisition system recorded also
a stopped-$\pi$ trigger timing
defined by beamline counters
as BC1 $\otimes$ BC2 $\otimes$ BC3 $\otimes$ 
$\overline{\mbox{BC4}}$,
where the pulse-height threshold of BC2 was carefully set
in order to select only stopped-pion events which deposited more energy
than an electron, a muon or an in-flight pion event.
The relative timing of this triggering system to the TES events
was reconstructed during the offline analysis. 

Precise absolute energy calibration
was crucial for this measurement.
Because independent energy calibration was necessary for each of the 240 pixels,
it was essential to supply intense calibration x-rays
to achieve sufficient statistics for $in$-$situ$ calibration.
An x-ray tube source was installed
as shown in Fig. \ref{fig_setup}.
Characteristic x-rays were produced by shining
a controllable flux of x-rays,
which was generated by an electron gun with a Rh target,
on 99.999\% pure chromium and cobalt pieces.
These calibration x-rays traveled through the hollow, 
conical carbon target to the TES spectrometers.

\begin{figure}[hbtp]
\centering\includegraphics[width=0.65\linewidth]{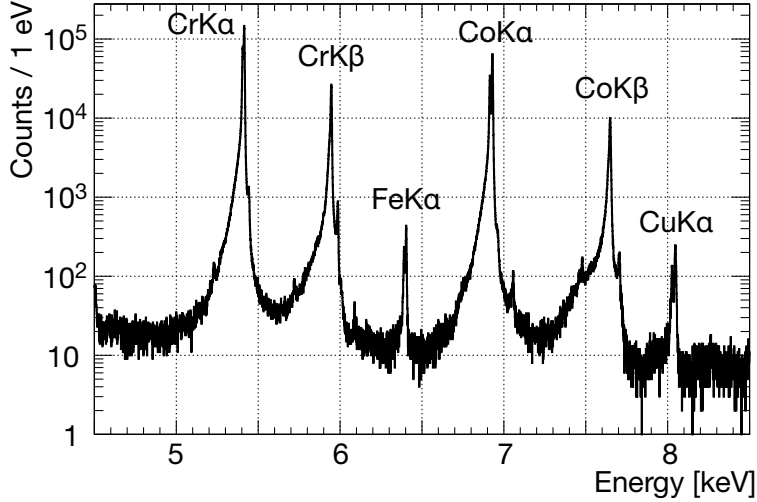}
 \caption{An x-ray energy spectrum obtained from a tube-source x-ray generator
 shining on Cr and Co calibration pieces without a pion beam.
 The Cr $K_\alpha$, Cr
 $K_\beta$, Co $K_\alpha$, and Co $K_\beta$ lines are used for the
 energy calibration. 
 Lower-yield x-rays, Fe $K_\alpha$ (6.4 keV) and Cu $K_\alpha$ (8.0 keV),
 originate respectively from the stainless steel vacuum fittings
 of the tube source and the metal structure of the 50 mK detector package.}
\label{fig_calib}
\end{figure}

Figure \ref{fig_calib} shows
a summed x-ray energy spectrum of 209 working TES pixels
obtained with the x-ray generator without a pion beam.
The achieved energy resolution was 4.6 eV (FWHM) at 6.4 keV
with a count rate of 4.4 Hz/pixel.
The energy calibration was performed
with the four calibration x-ray lines,
Cr $K_\alpha$, Cr $K_\beta$, Co $K_\alpha$, and Co $K_\beta$, 
with natural cubic-spline interpolation.
Energies and natural widths of the calibration x-rays
were fixed with the reference values \cite{Holzer1997}
in the spectral fitting.
Lower-yield x-rays, Fe $K_\alpha$ (6.4 keV) and Cu $K_\alpha$ (8.0 keV),
originated respectively from stainless steel vacuum fittings in the tube source
and Cu inside of the detector package. These were not used for calibration.

\section{Results}

\begin{figure}[hbtp]
\centering\includegraphics[width=0.85\linewidth]{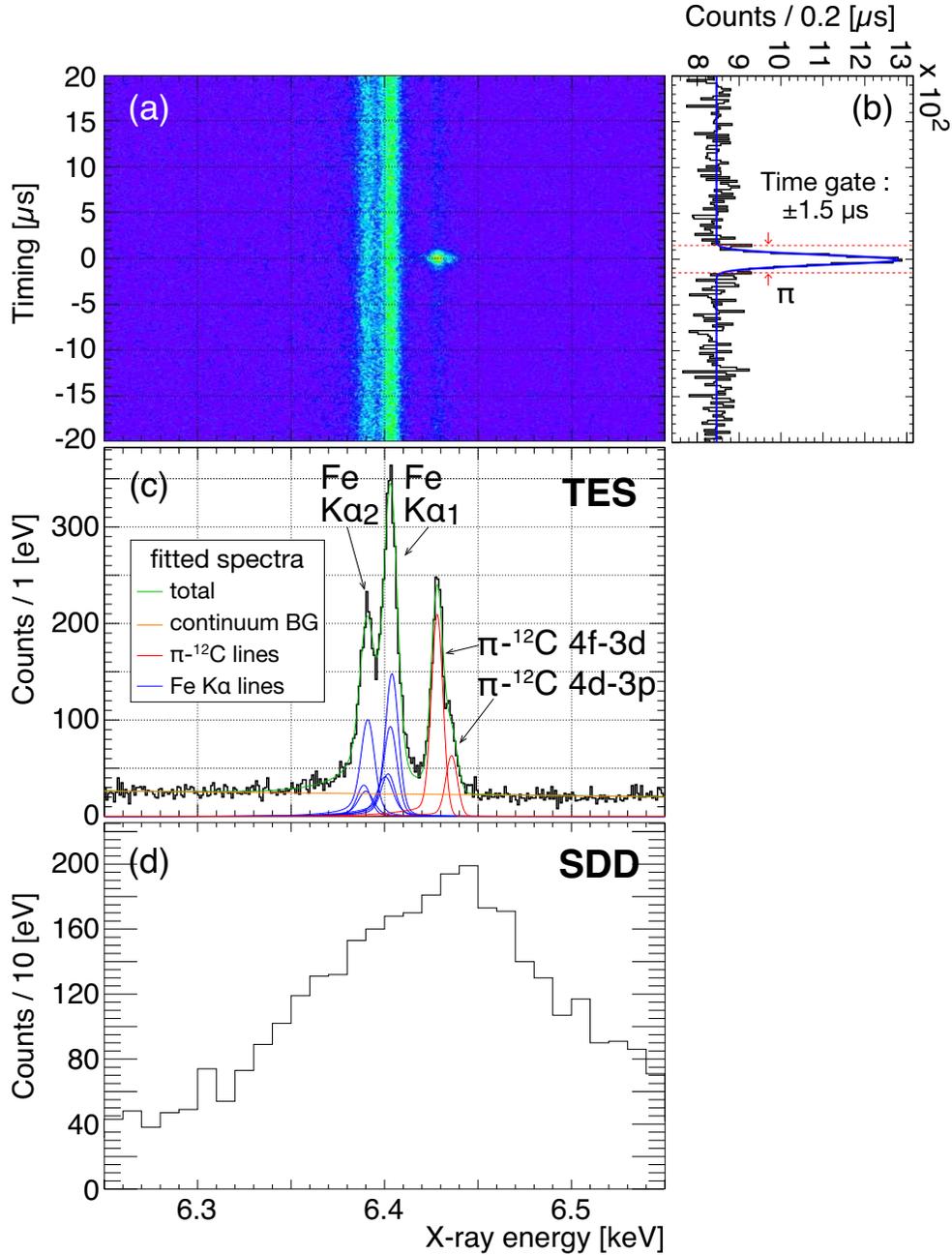}
\caption{Results of the measured x-ray time
 and energy distributions for stopped-$\pi^-$ trigger events.
 (a) A correlation plot of
 the time difference between pion arrival and x-ray detection vs
 the x-ray energy measured by the TES array.
 (b) The projection on the time axis
 showing timing resolution of 1.2 $\mu$s (FWHM).
 A time gate of $\pm$ 1.5 $\mu$s is used in the analysis.
 (c) The projection on the energy axis
  by selecting stopped-$\pi^-$ time gate indicated in (b), 
  where the fitted components for Fe $K_\alpha$ and 
 $\pi$-$^{12}$C x-rays
  are shown as well.
 (d) The same spectrum measured by the reference SDD
  having a FWHM energy resolution of $\sim$165 eV.}
\label{fig_result}
\end{figure}

Time and energy distributions of $\pi$-$^{12}$C x-rays
measured with the 209 working TES pixels
are shown in Fig. \ref{fig_result}.
The data were accumulated for 13.5 hours
with a pion-beam intensity of 1.45 MHz 
and a stopped-$\pi$ trigger rate of 34.5 kHz.
Figure \ref{fig_result} (b) shows a distribution of the time difference
between pion arrival and x-ray detection with the TES array
for stopped-$\pi^-$ trigger events.
A clear peak was observed with a time resolution of 1.2 $\mu$sec FWHM.
Figure \ref{fig_result} (a) shows a correlation plot of the time difference
versus the x-ray energy measured by TESs.
At around 6.43 keV, a clear time-energy correlation
corresponding to $\pi$-$^{12}$C $4f \to 3d$ x-ray was observed.
Figure \ref{fig_result} (c) shows an x-ray energy spectrum
of only those events recorded within the
stopped-$\pi^-$ timing gate ($\pm$1.5 $\mu$sec)
that is indicated in Fig. \ref{fig_result} (b).
A sharp peak from the $\pi$-$^{12}$C $4f \to 3d$ transition
is observed.
The TES spectrometer has sufficient energy resolution to observe
the parallel transition $4d \to 3p$.
Figure \ref{fig_result} (c) shows contributions from 
those $4f \to 3d$ and $4d \to 3p$ lines whose centroids are only 7.4 eV apart.
The timing selection improved the peak-to-background ratio
of the $\pi$-$^{12}$C line from 2 to 10.
As a reference in this experiment,
a SDD is installed just beneath the carbon target
as shown in right panel of Fig. \ref{fig_setup}.
The FWHM energy resolution of the SDD of $\sim$ 165 eV
is not nearly enough to resolve any of
the important features of the spectrum (Fig \ref{fig_result} (d)).

The tube source produces Cr and Co calibration x-rays constantly
during data acquisition with the pion beam.
The energy calibration with Cr and Co x-rays is recalculated
every 2 hours to mitigate gain drift.
A fit of the energy spectrum with the in-beam condition of $\pi^-$ beam
gives a FWHM energy resolution
of 6.8 eV at 6.4 keV
with a TES hit rate of 4.8 Hz/pixel, of which 0.4 Hz is
due to the pion beam.
The deterioration of the energy resolution
and the size of low-energy and high-energy tail components
in the non-Gaussian energy response of TESs
both increase as a function of pion beam intensity.
These beam-correlated effects are explained
by the production of thermal cross-talk pulses
which are due to direct charged-particle hits
in the silicon substrate of the TES array.
The detail of the detector response and the absolute energy calibration was
presented in a separate paper \cite{LTD16}.

The x-ray generator fluoresces materials
containing iron (e.g., the stainless steel vacuum fittings around the target cell);
therefore characteristic x-rays of iron
are observed that are uncorrelated with the pion-beam timing.
The Fe $K_{\alpha1}$ (6.404 keV) 
and Fe $K_{\alpha2}$ (6.391 keV) lines
were used to evaluate the accuracy of energy calibration
that was then used to determine $\pi$-$^{12}$C x-ray energy.
A result of a spectral fit of Fe $K_\alpha$ and $\pi$-$^{12}$C x-rays 
is shown in Fig. \ref{fig_result} (c).
The energy-calibration accuracy is assessed
by the fit to the measured Fe $K_{\alpha}$ line.
The measured energy of our Fe $K_{\alpha11}$ line is:
\begin{eqnarray*}
  6404.07 \pm 0.10 \mbox{(stat.)} ^{+0.06}_{-0.04} \mbox{(syst.) eV}
\end{eqnarray*}
where the first error is statistical and the second is systematic.
The quoted systematic uncertainty is a quadratic summation
of the contributions from continuum background parameter 
and asymmetry of the fit function.
A comparison with the reference value
of 6404.148(2) eV \cite{Holzer1997}
for pure, metallic iron shows good agreement within the errors.
To determine the energy of pionic carbon transition x-ray,
the critical issue is the uncertainty precisely at the lines of interest, 6430 eV. 
It is difficult to assess the systematic uncertainty
introduced by our choice for the calibration curves in its full generality.
Fortunately, the Fe $K_{\alpha}$ line at 6404 eV is very close to the line of interest,
and this good agreement with the reference value validated the choice
of the present energy calibration curves.

Fits to our energy spectra determined
the energies of the
$\pi$-$^{12}$C $4f \to 3d$ and $4d \to 3p$ transition x-rays
and their yield ratios to be:
\begin{eqnarray*}
E (4f \to 3d) &=& 6428.39 \pm 0.13 \mbox{(stat.)} \pm 0.09 \mbox{(syst.) eV} \\
E (4d \to 3p) &=& 6435.76 \pm 0.30 \mbox{(stat.)} ^{+0.11}_{-0.07}  \mbox{(syst.) eV} \\
I(4d \to 3p) / I(4f \to 3d) &=& 0.30 \pm 0.03 \mbox{(stat.)} \pm 0.02 \mbox{(syst.)}
\end{eqnarray*}
where the first error is statistical and the second is systematic.
The quoted systematic uncertainty is a quadratic summation
of the contributions from uncertainties of
the energy calibration,
the non-Gaussian response function,
and the timing window width.
The tail component of Fe $K_{\alpha}$ line affects the fit results.
We assessed it by varying the relative strength of the Fe x-ray tail
by changing the timing window. The Fe tail is the main source
of systematic errors both for x-ray energies and yield ratio.
The energy dependence of transmissions of these two $\pi$-$^{12}$C peaks
is negligible to determine the yield ratio,
since these energies are very close and no absorption edge structure
exists around those peaks.

\begin{table}[t]
 \caption{Calculated values of pionic $^{12}$C electromagnetic 
 energies and strong-interaction energy shifts 
 via the Seki-Masutani potential \cite{SM83}.
 For the $4f \to 3d$ and $4d \to 3p$ transitions,
 electron-screening effects are assessed 
 with the cases of filling one $1s$ electron 
 and two $1s$ electrons in the K-shell.
 The experimental results are shown as well.}
\label{table_calc}
\centering
\begin{tabular}{cccccccc}
 \hline
 \hline
 State & K.G. & 
 \multicolumn{2}{c}{Vacuum polarization} &
 Nuclear & Relativistic & Strong & Total \\
  & energy & 
 $\alpha(Z\alpha)$ & $\alpha^2(Z\alpha)$ &
 finite size & recoil effect & interaction & energy \\
  & (eV) & (eV) & (eV) &
 effect (eV) & (eV) & effect (eV) & (eV) \\
 \hline
 $3p$ & $-$14685.15 & $-$ 11.56 & $-$0.08   & $+$ 0.01     & $-$0.02 & $-$0.78      & $-$14697.58 \\
 $3d$ & $-$14682.65 & $-$   5.39 & $-$0.04   & $+$ 0.0005 & $-$0.02 & $<10^{-4}$ & $-$14688.10 \\
 $4d$ & $-$8259.04   & $-$   2.10 & $-$0.02   & $+$0.0003  & $-$0.01 & $<10^{-4}$ & $-$8261.17\\
 $4f$  & $-$8258.59   & $-$   0.72 & $-$0.004 & $+$0.0003  & $-$0.01 & $<10^{-4}$ & $-$8259.32\\
 \hline
 \hline
\end{tabular}
~\\
\vspace{5mm}   
\centering
\begin{tabular}{ccccl}
 \hline
 \hline
 Transitions &  \multicolumn{3}{c}{Electron screening effect (eV)} & Transition \\
   & Configuration & K-shell & L-shell & energy \\
   &  & contribution & contribution & (eV) \\
 \hline
             & no electron      & -       & -       & 6428.78 \\
  $4f \to 3d$  & $1s^1$ $2s^2$ $2p^1$ & $-$0.19 & $-$0.02 & 6428.57 \\
             & $1s^2$ $2s^2$ $2p^1$ & $-$0.31 & $-$0.01 & 6428.46 \\
\cline{2-5}
 & \multicolumn{3}{r}{Experimental result (this work) :} & 6428.39 $\pm$ 0.13 $\pm$ 0.09 \\
\hline					               
            & no electron      & -       & -       & 6436.41 \\
 $4d \to 3p$  & $1s^1$ $2s^2$ $2p^1$ & $-$0.25 & $-$0.02 & 6436.14 \\
            & $1s^2$ $2s^2$ $2p^1$ & $-$0.42 & $-$0.01 & 6435.98 \\
\cline{2-5}
 & \multicolumn{3}{r}{Experimental result (this work) :} & 6435.76 $\pm$ 0.30 $^{+0.11}_{-0.07}$ \\
 \hline
 \hline
\end{tabular}
\end{table}

We have calculated the $3p$, $3d$, $4d$ and $4f$ energy levels
of the pionic $^{12}$C
using only the electromagnetic (EM) interaction
and tabulated the results in Table \ref{table_calc}.
These EM values are calculated from the Klein-Gordon equation
including vacuum polarization with higher-order correction,
the relativistic-recoil effect and the nuclear-finite-size effect.
The latest charged pion mass, 139.57018(35) MeV/$c^2$,
given by the particle data group \cite{PDG} is used.

Strong-interaction effects on these energy levels
were assessed via the Seki-Masutani potential \cite{SM83}.
The strong-interaction shift of the $3p$ level from its EM value
is calculated to be 0.78 eV,
while the shifts of the $4f$, $4d$, $3d$ levels are below $10^{-4}$ eV.

The effect of the electrons populating the K- and L-shells,
the so-called electron-screening effect, is not negligible 
especially in the solid target.
The electron-screening correction to the x-ray energy is largely
determined by the number of $1s$ electrons in the atom, which depends on
the balance between Auger-electron emission and the electron-refilling process
from neighboring atoms.
We have estimated the electron screening effect
with the ``$Z-1$'' approximation, namely the captured pion 
screens one unit of the nuclear charge seen by the electrons.
Thus, the maximum electron-screening correction is given 
by the $1s^2$ $2s^2$ $2p^1$ electronic configuration.
In Table \ref{table_calc},
the EM values of the $4f \to 3d$ and $4d \to 3p$ transitions
are tabulated for the no electron,
$1s^1$ $2s^2$ $2p^1$ and $1s^2$ $2s^2$ $2p^1$ electronic configurations.
Here, the electron-density functions in each configuration are evaluated 
using the hydrogenic wave function with effective nuclear charge
based on Hartree-Fock calculations.
The experimental results are consistent 
with the $1s^2$ $2s^2$ $2p^1$ configuration within the errors.

\section{Conclusion}

We observed the $\pi$-$^{12}$C $4f \to 3d$ transition x-ray line
with a novel 240-pixel microcalorimetric x-ray detector
based on transition edge sensors.
The achieved averaged energy resolution is
6.8 eV FWHM at 6.4 keV under a high-rate pion beam intensity of 1.45 MHz.
The timing resolution is 1.2 $\mu$sec FWHM.
Absolute energy calibration is realized
by an x-ray generator shining on calibration metals during the data acquisition.
The resulting systematic uncertainty in the $\pi$-$^{12}$C $4f \to 3d$ transition x-ray energy
is less than 0.1 eV, 
which meets our goal for a future measurement
of the kaonic-helium $3d \to 2p$ x-ray energy.

The TES spectrometer had sufficient energy resolution to observe
the parallel transition $\pi$-$^{12}$C $4d \to 3p$ x-ray for the first time.
The strong-interaction effect of the $3p$ level is not negligible
because it has smaller angular momentum than the $3d$ level.
The measured x-ray energy of the parallel transition $4d \to 3p$
was found to be consistent with the calculated strong-interaction effect
assessed via the Seki-Masutani potential \cite{SM83}.

Our data allow the determination of
the electron population status of the atoms.
Both the $4f \to 3d$ and $4d \to 3p$ transition energies obtained
favor two $1s$ electrons in the K-shell.
Moreover we have determined the yield ratio
between the $4f \to 3d$ and $4d \to 3p$ transitions.
Whether the observed x-ray yields can be consistently explained
by a cascade calculation with the electronic configuration
of two filled $1s$ electrons in the K-shell
remains to be solved in future theoretical work.

We successfully demonstrated the feasibility of hadronic-atom x-ray spectroscopy
with an absolute energy uncertainty of 0.1 eV in the 6 keV energy region.
This is an important milestone towards a more general use of high-resolution
microcalorimeter spectrometers at charged-particle beamlines.

\ack
The authors thank
K. Deiters and the PSI staff for beamline coordination and operation.
J. Uhlig thanks the Knut and Alice Wallenberg Foundation for their continued support. 
This work was partly supported by RIKEN, NIST, KEK,
a Grants-in-Aid for Scientific Research 
from MEXT and JSPS (Nos. 25105514, 26707014, 24105003,
15H05438 and 15H00785), 
the strategic young researcher overseas visits program for
accelerating brain circulation by JSPS (No. R2509),
Incentive Research Grant from RIKEN,
Mitsubishi Foundation (26145),
and the NIST Innovations in Measurement Science Program.

\end{document}